\documentclass[aps,superscriptaddress,amsmath,amssymb,showpacs,twocolumn]{revtex4-1}
\usepackage{graphicx}

\usepackage[utf8]{inputenc}
\newcommand{\beq}{\begin{equation}}
\newcommand{\eeq}{\end{equation}}
\newcommand{\beqa}{\begin{eqnarray}}
\newcommand{\eeqa}{\end{eqnarray}}
\newcommand{\bsub}{\begin{subequations}}
\newcommand{\esub}{\end{subequations}}

\newcommand{\rem}[1]{}
\newcommand{\refe}[1]{Eq.~(\ref{#1})}

\begin{document}
\title{Tunable spin-polaron state in a singly clamped semiconducting carbon nanotube}
\author{F. Pistolesi}
\affiliation{
Univ. Bordeaux, LOMA, UMR 5798,  Talence, France.\\
CNRS, LOMA, UMR 5798, F-33400 Talence, France.\\
}

\author{R. Shekhter}
\affiliation{
University of Guthenberg, Sweden
}

\begin{abstract}
We consider a semiconducting carbon nanotube (CNT) laying on a ferromagnetic insulating substrate 
with one end depassing the substrate and suspended over a metallic gate. 
We assume that the polarised substrate induces an exchange interaction acting as a local magnetic 
field for the electrons in the non-suspended CNT side. 
Generalizing the approach of I.~Snyman and Yu.V.~Nazarov [Phys. Rev. Lett. {\bf 108}, 076805 (2012)] 
we show that one can generate electrostatically a tunable spin-polarized polaronic state 
localized at the bending end of the CNT.
We argue that at low temperatures manipulation and detection of the localised quantum spin state is possible.  
\end{abstract}

\date{\today}

\pacs{85.85.+j,71.38.-k, 85.75.-d} 


\maketitle

\newcommand{\Eex}{E_{\rm ex}}
\newcommand{\com}[1]{{\tt #1}}


%
Nanoelectromechanics with suspended carbon nanotubes evolved rapidly in last few 
years \cite{lassagne_coupling_2009,steele_strong_2009,meerwaldt_carbon_2012,moser_nanotube_2014,benyamini_real-space_2014,schneider_observation_2014,zhang_interplay_2014}.
Recently I. Snyman and Yu.V. Nazarov \cite{snyman_polarons_2012} considered 
a semiconducting CNT laying on an insulating substrate with one end of it suspended.
A metallic gate below both the insulating substrate and the suspended part of the CNT generates 
an homogeneous electric field (cf.~Fig.~1 of Ref. \cite{snyman_polarons_2012}).
The mechanical bending of the suspended part of the nanotube induces then a spatial inhomogeneity 
of the electrostatic potential along the CNT forming a minimum at the deformable end of the wire. 
The competition between such an electrostatic bending with both the elastic potential of the CNT and 
the quantum rigidity of the electronic wave function  
makes the mechanical bending as well as the formation of the localized polaronic state 
at the movable end of the CNT to occur as a first order phase transition as a function 
of the electric field.
The estimate for the critical field for a realistic experimental set up was predicted
in Ref. \cite{snyman_polarons_2012} to be 0.01 V/$\mu$m.

An impressive effort of the nanoelectronics community 
is currently deployed to manipulate and exploit the electronic spin 
degrees of freedom in transport devices (spintronics) \cite{zutic_spintronics:_2004}.
In this context the possibility of {\em magnetic gating}, {\em  i.e.} 
the use of ferromagnetic leads inducing magnetic exchange 
fields $\Eex/\mu_B$ (with $\mu_B$ the Bohr magneton) 
on the electronic spin is currently actively investigated \cite{pasupathy_kondo_2004,hamaya_kondo_2007,cottet_controlling_2006,feuillet-palma_conserved_2010}.
More surprisingly such exchange fields can have remarkable consequences also 
on the dynamics of a nano-mechanical system for which 
dynamical (shuttle) instabilities, strong spin-polarized currents, and cooling have been predicted \cite{shekhter_spintronic_2012,kulinich_single-electron_2014,stadler_ground-state_2014}

\begin{figure}[tbp]
\includegraphics[width=3.5in]{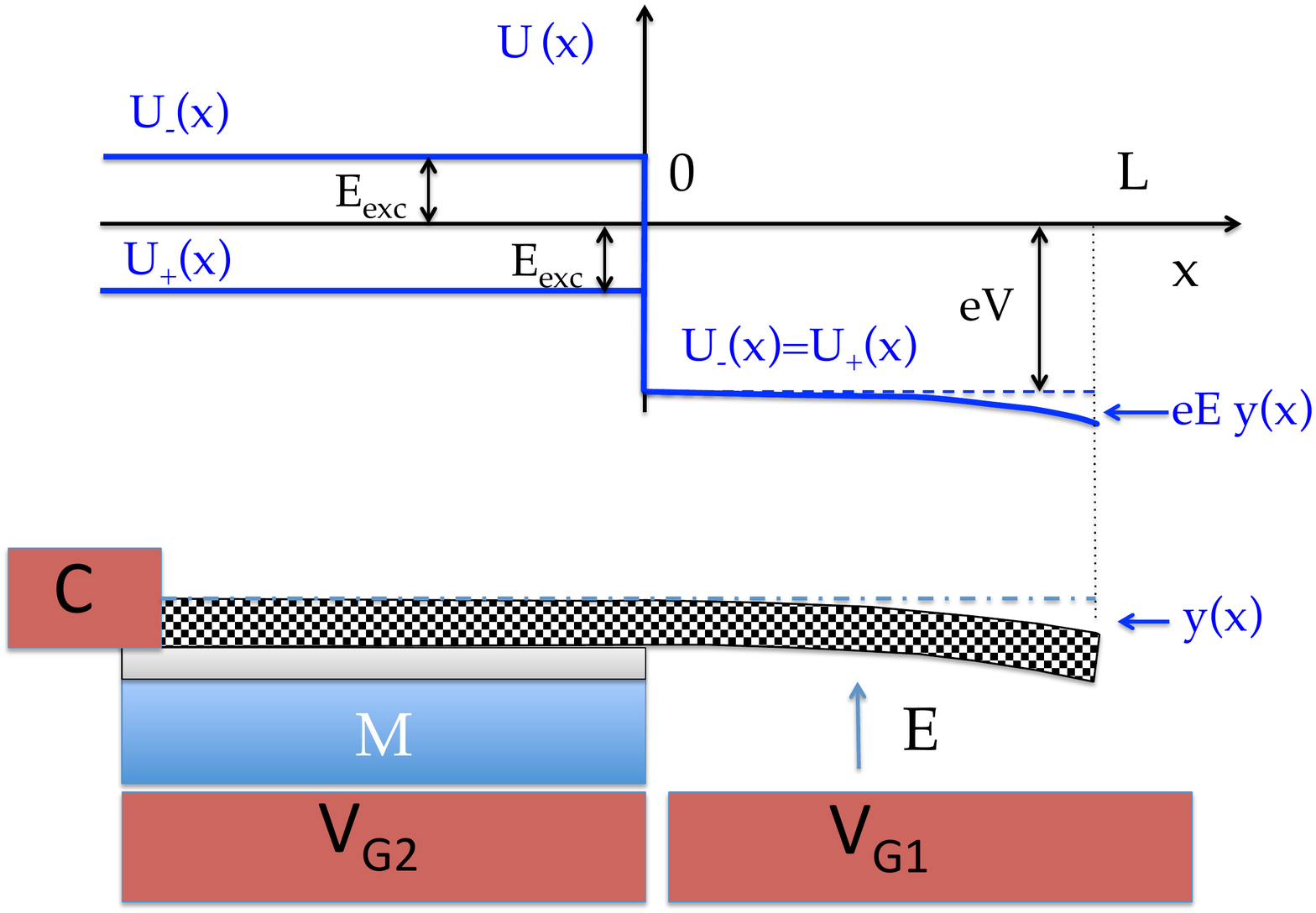}
  \caption{Schematic of the system considered: a CNT laying on a 
  magnetic substrate ($C$) and protruding out of a length $L$. 
  Two independently adjustable gates ($V_{G1}$ and $V_{G2}$) 
 are shown and a contact ($C$) on the substrate side.
 The potential for spin up and down ($U_+$ and $U_-$) is also sketched
  above.}
\label{fig:schematic}
\end{figure}

In this paper we show that the system discussed by Snyman and Nazarov \cite{snyman_polarons_2012}
in presence of a magnetic dielectric substrate allows the formation of a localized fully polarized polaronic state.
The experimentally observed exchange energy $\Eex$ 
(see Ref. \cite{feuillet-palma_conserved_2010,shekhter_spintronic_2012}) 
 turns out to be as large as tens of Kelvins, thus 
being of the same order of magnitude of the localization energy for an electron in a CNT on 
the scale of the micrometer. 
This allows for a high tunability of the polaronic state by means
of two electric gates, 
below the suspended and non-suspended part of the CNT (see Fig.~\ref{fig:schematic}). 
As a result a continous electrostatic tuning of the localization length and the bound state energy 
can be achieved, forming a stability diagram of spin-up and spin-down polaronic states.
Detection of the state of the system can be envisaged by use of a nearby single-electron transistor,
for which the CNT tip acts as a gate \cite{brenning_ultrasensitive_2006}.
Fully electric manipulation of the mechanical and electronic spin state of the CNT is thus possible in this system. 

%
%
%
%
%

{\em The system.} 
Following Ref.~\cite{snyman_polarons_2012} let us consider a CNT 
laying on a substrate with a suspended part protruding out of a length $L$ (see Fig. 1). 
In Ref. \cite{snyman_polarons_2012} it has been shown that the wavefunction $\psi(x)$ of the 
electrons in the valence band of the CNT can be describled by a standard one-dimensional
Schroedinger equation with an effective mass $m^*=0.6 m_e a_0/r$, where $m_e$ is the electronic
 mass, $a_0$ the Bohr radius, and $r$ the radius of the CNT.
The variable $x$ parametrizes the position along the CNT, its value is $0$ at the edge of the 
substrate and $L$ on the tip of the suspended part.
The length of the CNT on the substrate is supposed to be $\gg L$, and is 
taken infinity for simplicity. 
Then vanishing boundary conditions apply at $x=L$ and $x=-\infty$.
As in Ref. \cite{snyman_polarons_2012} the CNT can bend with a  
displacement $y(x)$ (for $0 < x < L$) in the direction orthogonal to the substrate.
The elastic energy cost reads $I Y \int dx  \left[y''(x) \right]^2/2$, where $I=6.4 \pi a_0 r^2 $ is 
the second moment of area of the tube cross section and $Y$ the CNT Young modulus 
(of the order of 1.2 TPa).
Single clamping implies that $y(0)=y'(0)=0$ and $y''(L)=y'''(L)=0$.
In this paper we will restrain to the classical description of the deflection.
The tip of the CNT on the substrate is in tunneling contact with a metal whose 
chemical potential can be tuned close to the valence band of the CNT
by adjusting the electric potential.
Up to now the description followed closely Ref.~\cite{snyman_polarons_2012}. 
We introduce now the main difference: We will assume 
that the substrate is a magnetic insulator that induces an exchange 
interaction term $-\Eex \int_{-\infty}^0 dx \sigma |\psi_\sigma(x)|^2$ for the electrons 
being in the CNT over the substrate ($x<0$).
The variable $\sigma$ indicates the spin projection in the $z$ direction. 
This creates a spin-dependent potential so that the spin-up electrons ($\sigma=+$)
are attracted in the $x<0$ region.
In order to tune the potential we assume that two different gates 
are present, one below the magnetic substrate and an other one under the 
suspended part. 
By changing independently the potentials on the two gates it is possible to modify the electrostatic
potential $V$ and the electric field $E$ acting on the electrons on the suspended part 
(taking the non-suspended region as a reference for the potential, cf. Fig.~\ref{fig:schematic}).
We can then write the full Hamiltonian for the problem as follows ($\theta_x$ is the Heaviside function):
\beqa
	H&=& \sum_\sigma \int_{-\infty}^{L} \!\!\!\!\!dx 
\left[ 
	{\hbar^2  \over 2 m^*} \left| \partial_x \psi_\sigma (x)\right |^2 
	+ {I Y \over 2 } \left(\partial_x^2 y(x) \right)^2
	\right.
	\nonumber\\
	&&\left.
		 -\left(
	 	\Eex \sigma  \theta_{-x}
	 	+ eV \theta_x
		- eE y(x)
	\right)		|\psi_\sigma(x)|^2
	\vphantom{{\hbar^2  \over 2 m^*}}
\right]
\,.
\label{Hamiltonian}
\eeqa
The first term in \refe{Hamiltonian} gives the quantum kinetic energy, the second 
the elastic energy, and the third is a sum of three parts: the exchange energy, 
the electrostatic potential and its variation induced by the deflection [$y(x)$] of the 
CNT.
In Ref.~\cite{snyman_polarons_2012}, for $V=0$ and for $\Eex=0$, 
it has been shown that it exists a critical value of the electric field $E_c$ for which 
the ground state is an electronic localized state on the CNT suspended part.
The formation of the bound state is a first order transition: the CNT starts to bend
only for $E>E_c$ and a metastable bound state exists for $E_{c1}<E<E_c$.
At $E=E_c$ the localization length is thus finite and typically much shorter than $L$. 
In order to have a tunable bound state it is necessary to have a smooth transition from
the delocalized to the localized state. 
This is actually the typical case in quantum mechanics, by decreasing the depth of a potential well
that allows a bound state one can delocalize progressively the wave function.
The bound-state radius then diverges at the threshold for its appearance. 
We will thus see that the presence of $V$ and $\Eex$ allows to create a spin-dependent 
tunable bound state, that is associated to a displacement of the CNT tip. 
%

\begin{figure}[tbp]
\includegraphics[width=3in]{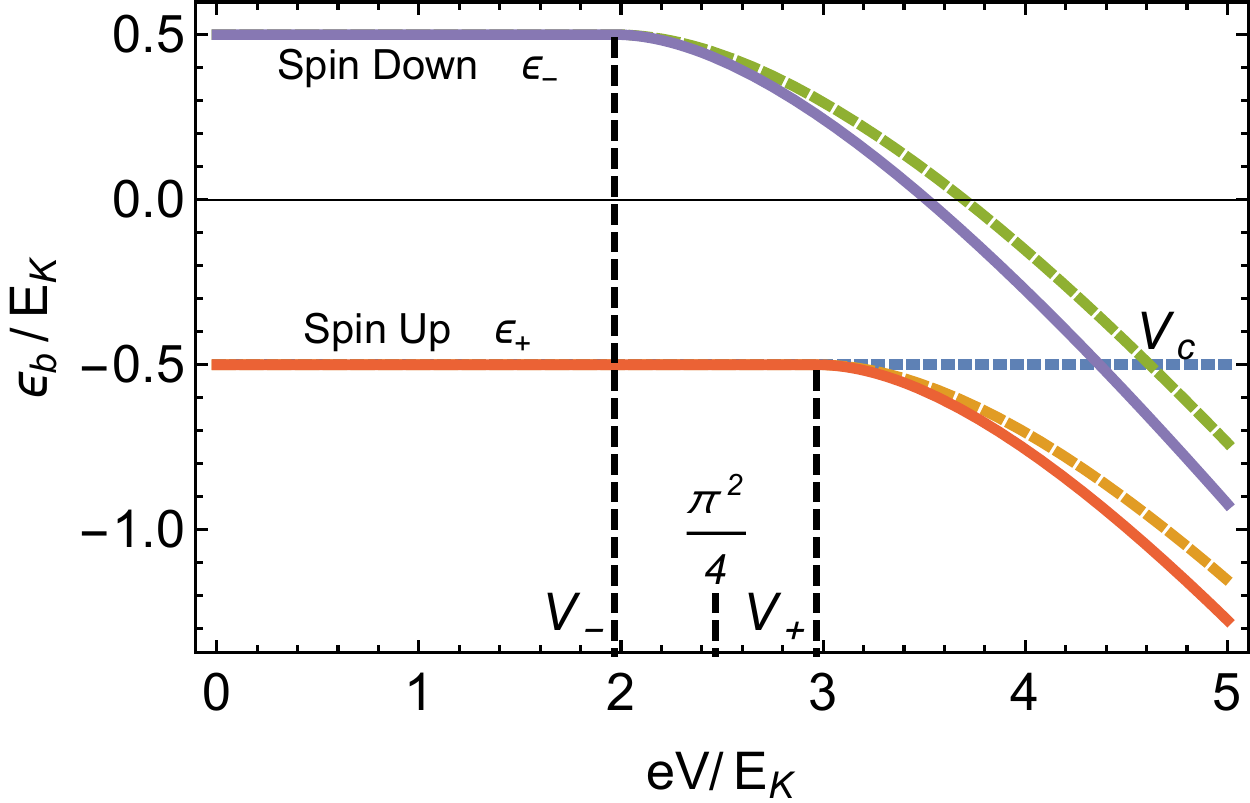}
  \caption{
  Dependence on the gate voltage $V$ of the two 
  spin-projection bound-state energy ($\epsilon_\sigma$) 
  for $\Eex/E_K=0.5$, $\alpha=0$ ({\em i.e.} $E=0$, dashed) and $\alpha=50$ 
  (full line). The bottom of the lower electronic band (spin-up) is shown dotted.
}
\label{fig:bound}
\end{figure}

{\em Electronic problem.} 
Let us begin with the purely electronic problem [$y(x)\equiv 0$ for all $x$].
The ground state can be found by solving the Schroedinger equation:
\beq
	\left[ -{\hbar^2 \partial_x^2 \over 2 m^*} 
	- \Eex \sigma  \theta_{-x} - eV \theta_x \right] \psi_\sigma(x) = \epsilon_\sigma \psi_\sigma(x) 
	\label{Hele}
	\,
\eeq
for each spin projection. 
The presence of a bound state is signaled by the existance of a solution of \refe{Hele}
with $\epsilon_\sigma<-\sigma \Eex$ the bottom of the relative band.
Taking $\epsilon_\sigma<-\sigma \Eex$ as a reference in energy the problem for each 
spin species reduces to that describled by \refe{Hele} with 
$\Eex \rightarrow 0$ and $eV \rightarrow eV-\sigma \Eex=U$. 
The solution can then be found by matching the wave function 
$\psi(x)=A e^{\kappa x}$ for  $x<0$ with 
$\psi(x)=B e^{i k x}+C e^{-i k x}$ for $x>0$ at 
$x=0$ asking the continuity of the wave function and of its derivative. 
The boundary conditions lead to the eigenvalue equation
$
	e^{-2 i k L} = - (ik+\kappa)/(ik-\kappa)
$
with 
$\kappa L = [-2m^* \epsilon_b/\hbar^2]^{1/2}$, 
$k L = [2m^* (U+\epsilon_b)/\hbar^2]^{1/2}$
and $\epsilon_b<0$ the bound state energy.
At the threshold $\epsilon_b\rightarrow 0^-$, thus there $\kappa$ 
vanishes and the eigenvalue equation reduces to $e^{-2ikL}=-1$.
This gives $kL=\pi/2$ and the threshold value for the potential 
$U_t = (\pi/2)^2 E_K $, with 
$ E_K={\hbar^2/(2 m^*L^2)} $ the kinetic energy scale.
For $U-U_t \ll E_K $  the localization length
$\kappa^{-1}= 2L E_K/(U-U_t)$ and diverges as anticipated.
By changing $U$  it is then possible to adjust the spread of the 
wave function on the magnetic substrate. 
Since the two spin species feel a different potential only on the substrate,
this allows to change continuously the energy difference of the up and 
down bound states.
The bound state energy for each spin state reads:
\beq
	\epsilon_\sigma =  -\sigma \Eex +\epsilon_b(eV-\sigma \Eex)
\eeq
with the threshold value for $V$: $ eV_\sigma = U_t+\sigma \Eex$.
A typical picture of the $eV$ dependence of the two bound states for 
$E=0$ is shown dashed in Fig. \ref{fig:bound}.
%
%
%
%
%
For  $V_-<V<V_+$ a unique bound state exists for spin down.
Let's define $V_c$ the value for which the down-spin energy 
crosses the up-spin bottom band: $\epsilon_-(V=V_c)=-\Eex$.
For $V_+ <V<V_c$ two bound states exist, but 
only the lowest one (spin up) is stable, since the spin-down lays above 
the bottom of the spin-up band, and any spin-flip perturbation 
allows its decay into the spin-up continuum.
Finally for $V > V_c$ two {\em stable}  bound states exist.
Their energy splitting has a maximum at $V_c$ and then monotonically 
decreases as a function of $V$. 
This is due to the reduction of the localization length reducing the effect of the 
exchange interaction that acts only for $x<0$.
Although both spin-up and spin-down polaronic states are stable at $V>V_c$ only one of them 
can be occupied due to the Coulomb blockade, whose repulsion energy turns out to be
much larger than the polaronic bound state energy ($\sim E_K$) at $L\gg$ 1 nm.
This fact allows the formation of a controllable single-electron 
fully spin polarized state at the protruding part of the CNT.


{\em Nanomechanical effects}.
We now consider how the system behaves when we let the CNT 
bend.
It is not possible any more to find the ground state energy analytically,
we will thus follow closely the variational method used in 
Ref.~\cite{snyman_polarons_2012} to which we refer for more details. 
We introduce the dimensionless variables
$z=x/L$, $h=H/E_K$, $f= y Y I/e E L^3$, $\phi_\sigma=\psi_\sigma \sqrt{L}$, and the 
coupling parameter $\alpha=(eE)^2 L^3/(Y I E_K)$. 
The problem can then be completely determined by giving only three independent coupling 
parameters: $\alpha$, $\mu=\Eex/E_K$, and $\nu=eV/E_K$.
The functional to be minimized reads:
\beq
	h = 
	\int_{-\infty}^1 \! \! \! \! dz 
	\left[ 
	{\phi_\sigma''}^2+\alpha \left( f \phi_\sigma^2 + {{f''}^2\over 2} \right) 
	-\left(
		\mu \sigma \theta_{-z}+\nu \theta_z
	\right) \phi_\sigma^2
	\right]
	\label{Hadim}
		\,.
\eeq
By writing $\phi(z)=\sum_{n=1}^M a_n (1-z)^n$ for $z>0$ and  $\sum_{n=1}^M a_n z^n e^{\kappa z}$ for $z<0$,
and $f(z)=\sum_{n=1}^M b_n z^{n+1}$ one can enforce the boundary conditions and minimize 
numerically the functional in order to find the parameters $\{a_n,b_n,\kappa\}$ and thus 
the ground state energy $\epsilon_\sigma$ with explicit expressions for the bending and the 
wavefunction.
The charge accumulated on the suspended part of the CNT in presence of an electric 
field induces a force that bends the tip of the CNT. 
The effective electronic potential  deepens and bending lowers
the bound state energy.
In particular it favors a stronger localization of the wave function on the tip (measured by $\kappa^{-1}$). 
 For $\Eex=eV=0$ in Ref.~\cite{snyman_polarons_2012} it is shown that 
the bound state forms with a first order phase transition for $\alpha>\alpha_c=312.03$.
In order to keep a smooth transition we will consider the case $\alpha<\alpha_c$ and 
investigate the dependence of the bound state energy and wavefunction on 
$eV/E_K$ for given values of $\Eex/E_K$.

Before considering the results of the numerical calculation it 
is useful to estimate analytically the typical range of the displacement of the 
CNT tip induced by the localization of the charge. 
Let us assume that the fraction $n<1$ of an electron charge 
is accumulated on the CNT tip uniformely.
A simple ansatz for the displacement is $f(z)= a z^2$.
It satisfies both the boundary conditions and the Euler equation $f''''=0$. 
Substituing it in \refe{Hadim} one has for the part proportional to $\alpha$:
\beq
	h_\alpha= \alpha (2 a^2+n {a\over 3})
	\label{energy}
\eeq
this functional has a minimum at  $a=-n/12=f(1)$. 
It gives a rough estimate of the dimensionless displacement of the tip 
by taking into account only the competition 
between the electric field and the elastic stiffness.
The effect of the other two parameters is hidden into the value of 
$n$, that cannot be larger than 1.

%
%
%
%
%
\begin{figure}[tbp]
\begin{tabular}{cc}
\includegraphics[height=1.in]{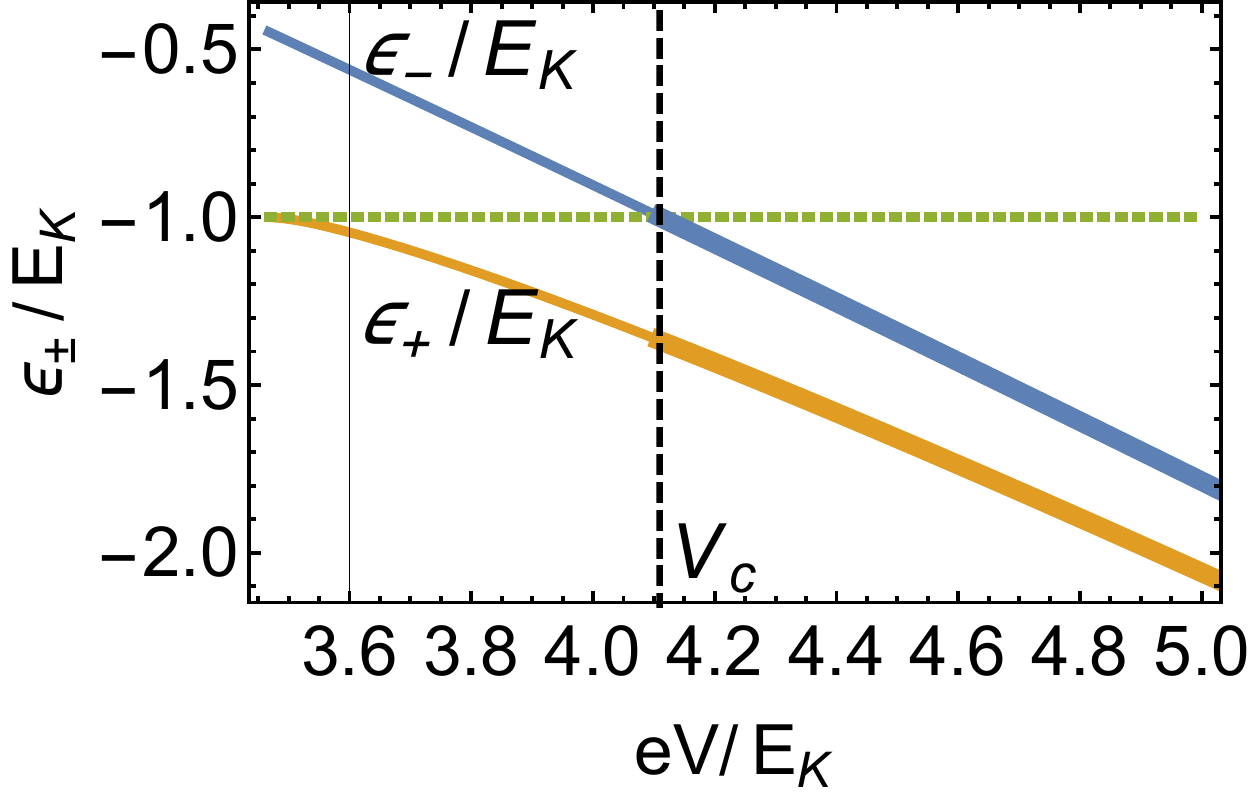}
&\includegraphics[height=1.in]{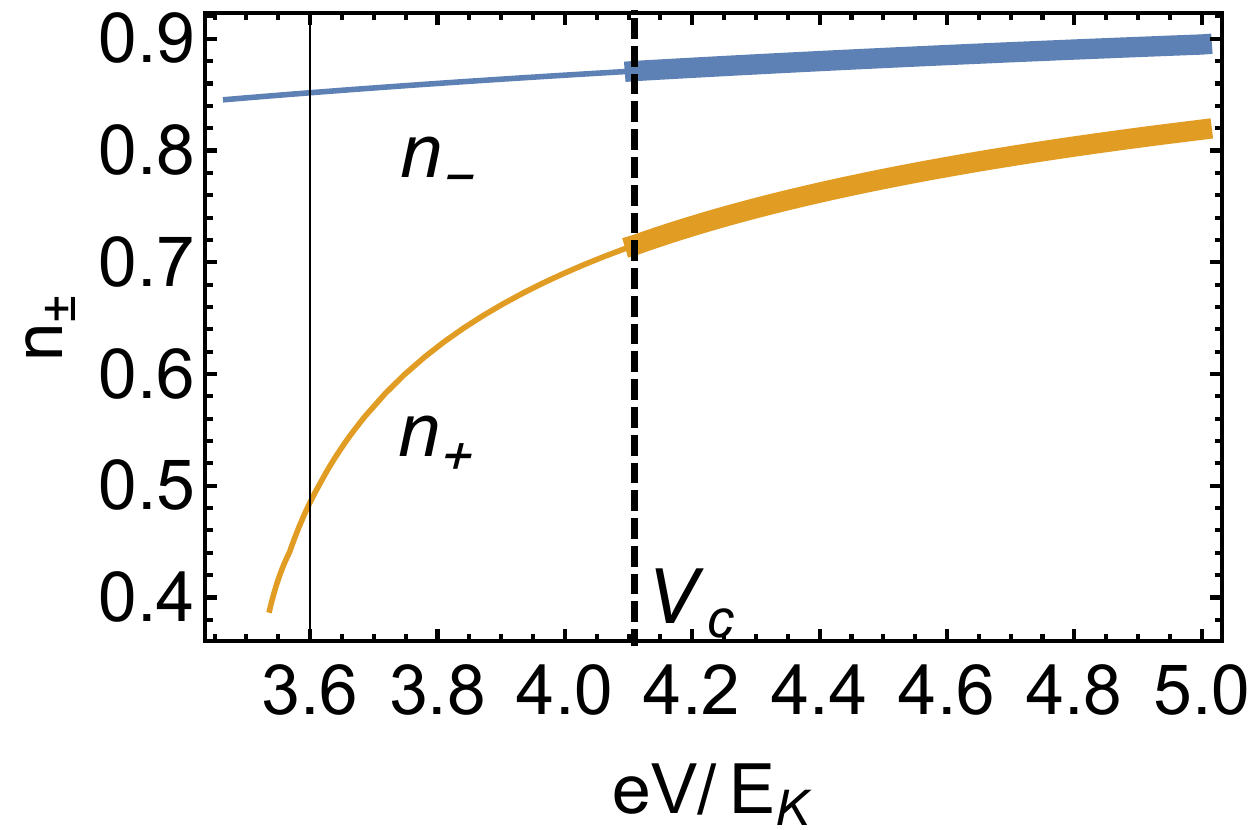} \\
\includegraphics[height=1.in]{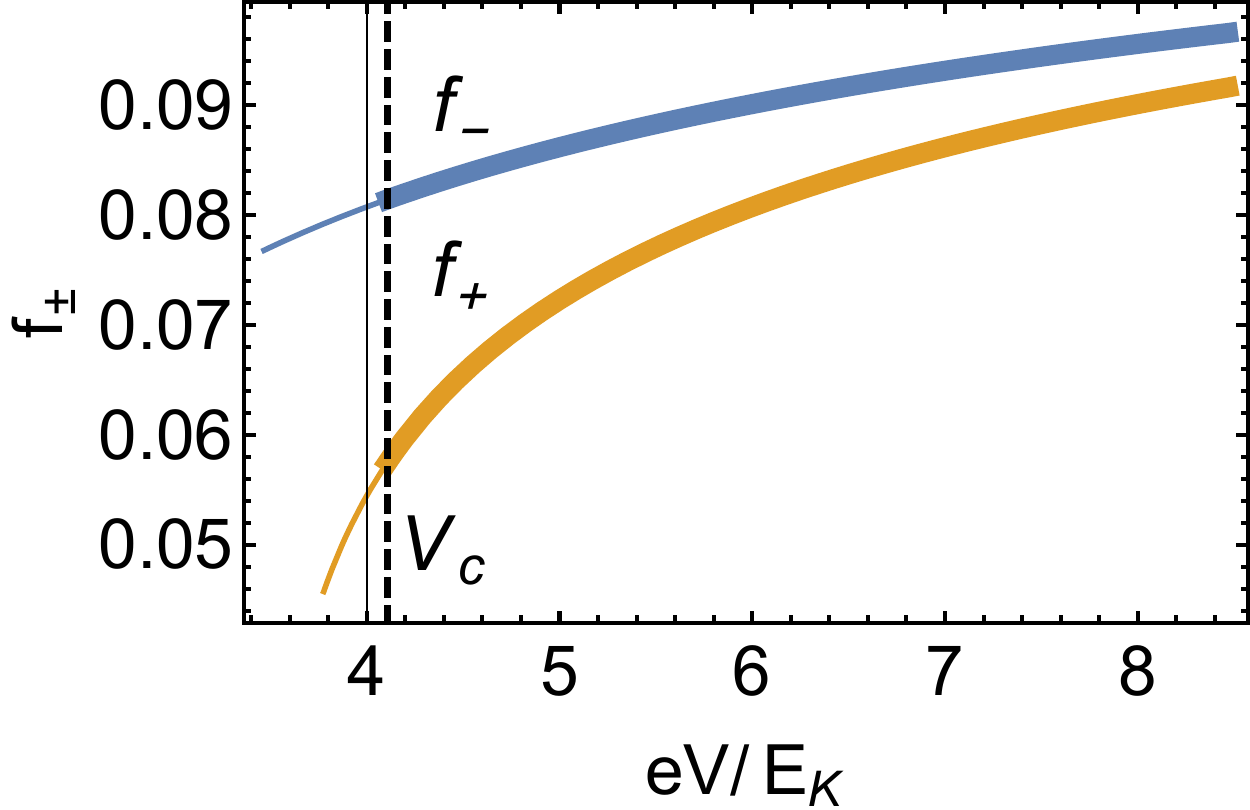}
&\includegraphics[height=1.in]{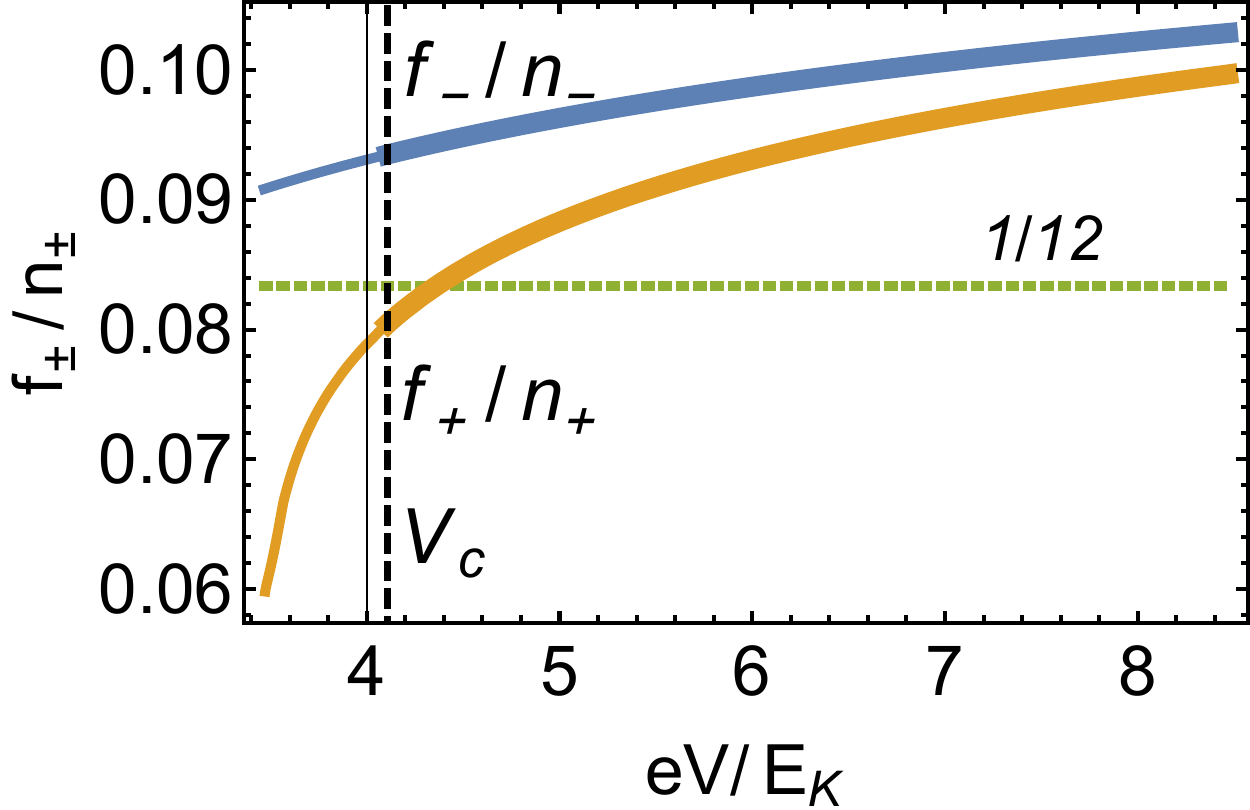} 
\end{tabular}
  \caption{
  For $\Eex=E_K$ and $\alpha=175$ gate voltage dependence 
  of the bound state energies (top-left), fraction of localized charge or spin
  (top-right), deflection of the tip (bottom-left), and ratio of deflection to localized charge 
  (bottom-right).}
\label{fig:gate}
\end{figure}
%
%

We present on Fig. \ref{fig:gate} the numerical results for $\alpha=175$ and 
$\mu=1$.  
One can see that the energy splitting of the two spin states is of the 
order of $E_K=\Eex$ (top-left panel).
Defining $n_\sigma = \int_0^1 dz \phi_\sigma^2$ the fraction of charge (and spin) 
localized on the suspended part of the CNT one finds that for $V=V_c$
both bound states present a finite value of $n_\sigma$ and 
$n_--n_+\approx 0.17$.
The difference is slowly reduced for larger values of the gate voltage.
The same can be said for the deflection of the tip of the CNT ($f_\sigma=f(1)$
for each spin state, bottom-left panel).
Finally the bottom-right panel shows that the ratio $f_\pm/n_\pm$ is actually
close to the rough estimate 1/12.
%
%
%
%
%
\begin{figure}[tbp]
\begin{tabular}{cc}
\includegraphics[height=1.in]{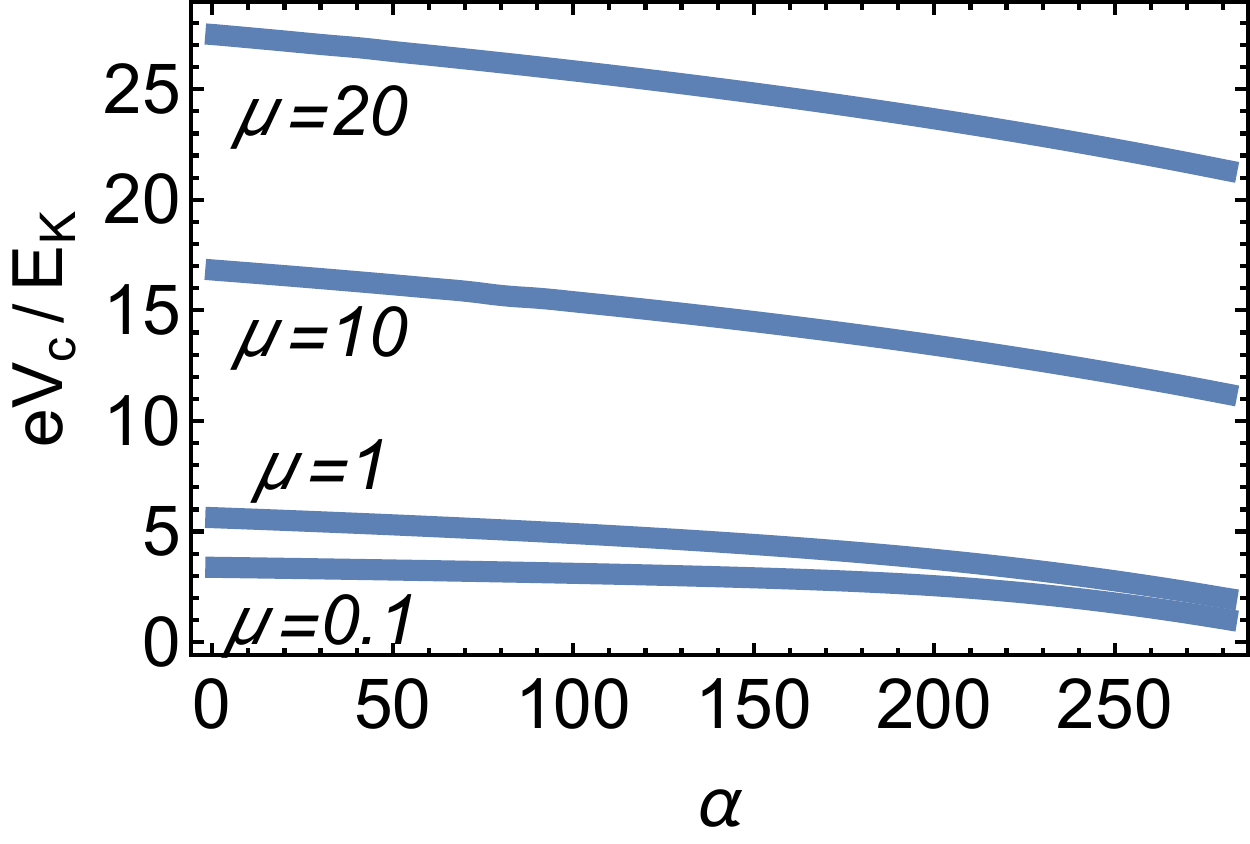} 
&\includegraphics[height=1.in]{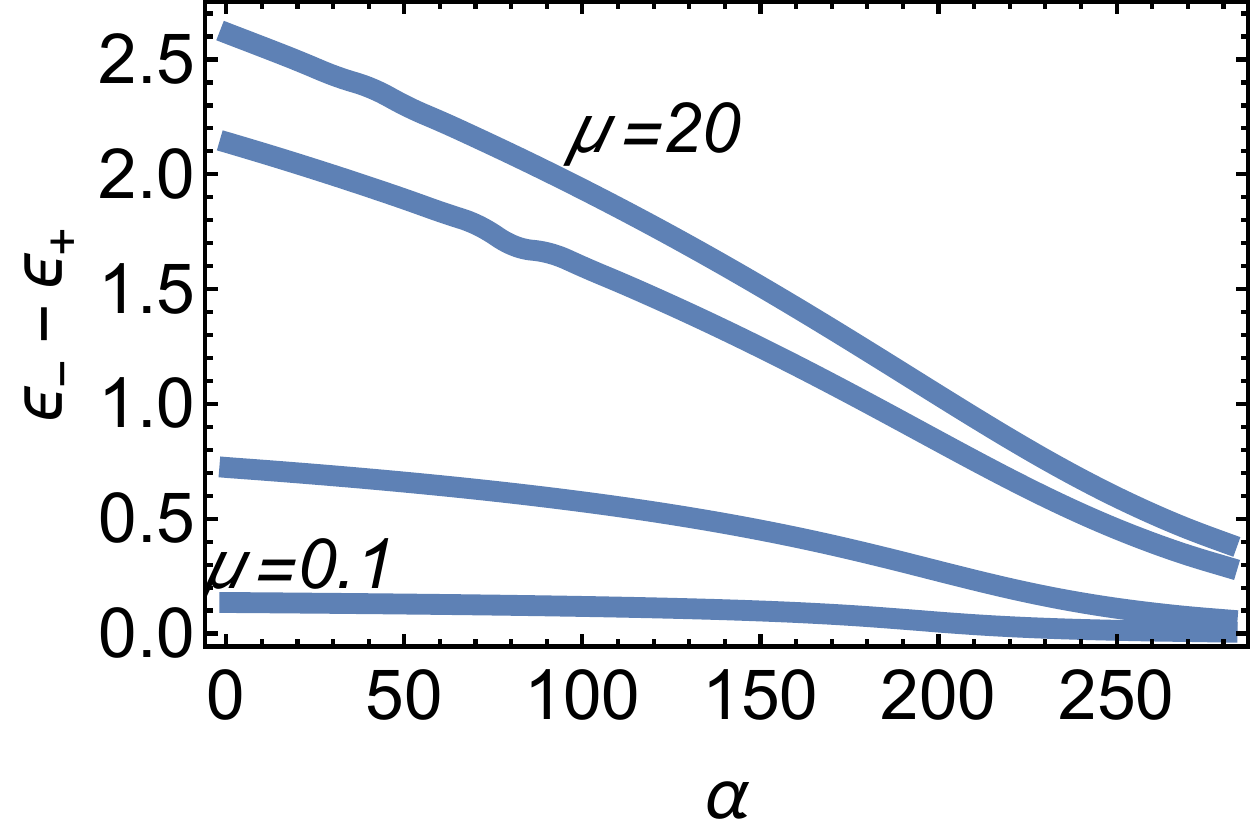} \\
\includegraphics[height=1.05in]{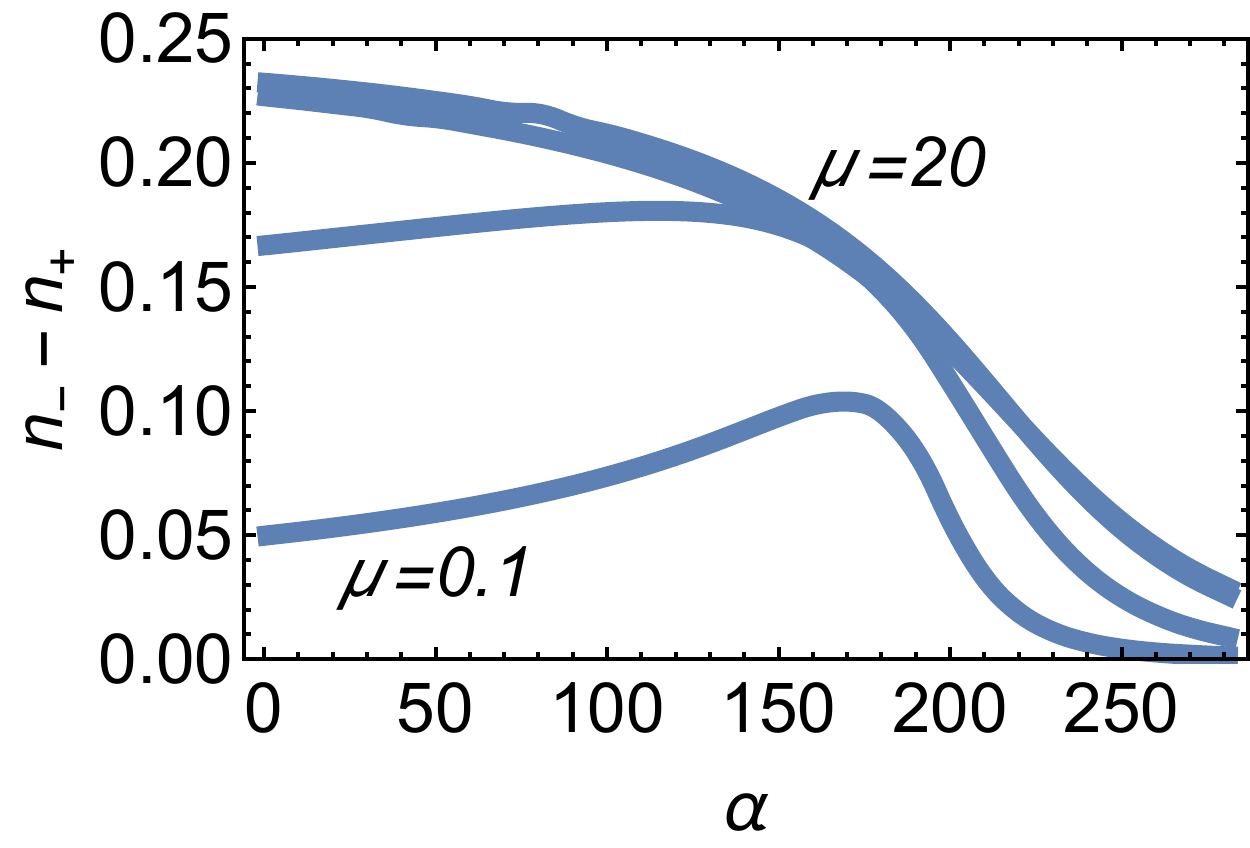}
&\includegraphics[height=1.in]{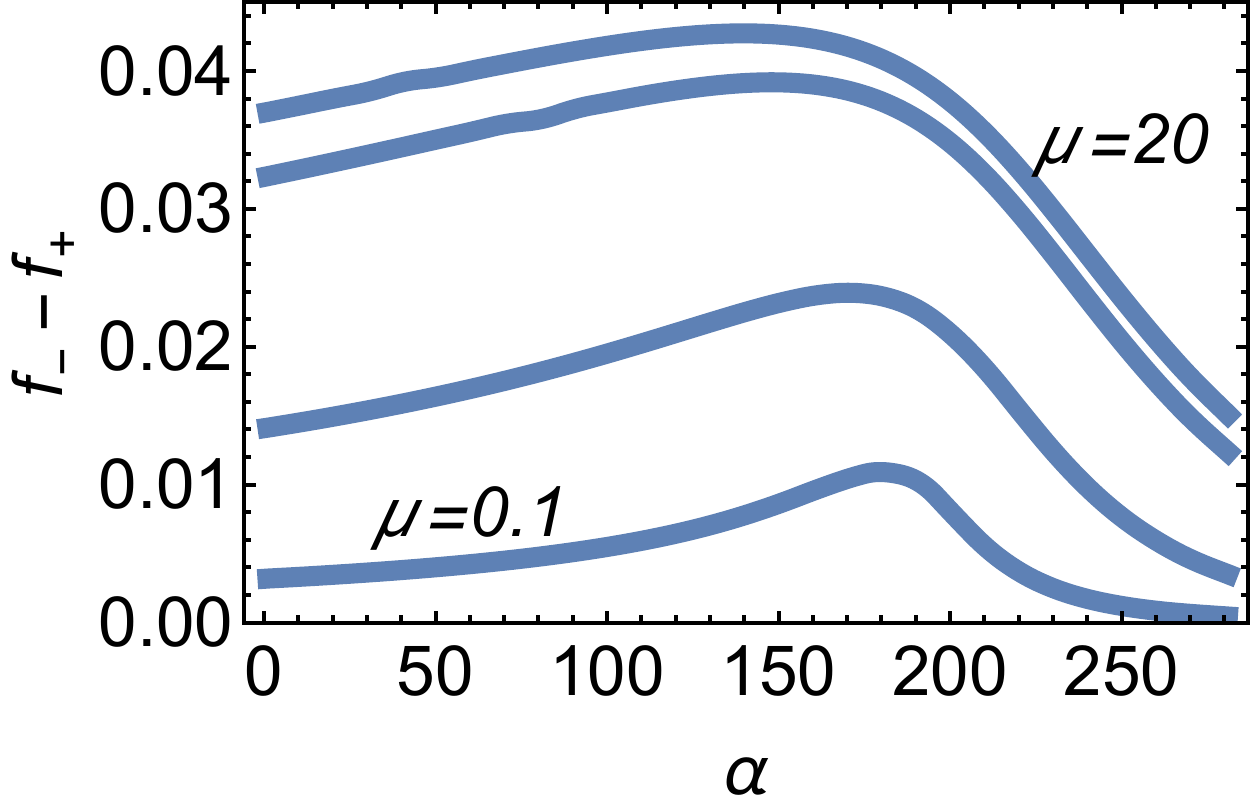}
\end{tabular}
  \caption{The four panels show as a function of $\alpha$ for $\mu=0.1, 1., 10, 20$
  the critical gate voltage value $V_c$ (top left), the energy splitting of the two bound states 
   $\epsilon_--\epsilon_+$ (top right), the difference in the fraction of localized charge 
   $n_+-n_-$ (bottom left), and the difference of the CNT tip deflection $f_--f_+$ (bottom-right);
   the last three quantities calculated at $V=V_c$. 
   }
\label{fig:critical}
\end{figure}
%
%

The plots of Fig.~\ref{fig:gate} show that a particularly important quantity is the value
of the physical parameters 
($\epsilon_\sigma$, $n_\sigma$ and $f_\sigma$) at the threshold $V_c$.
The dependence on $V$ is always monotonic and the maximum or minimum values are observed 
at $V_c$. 
In view of manipulating the spin state, the value at $V_c$ gives thus a very good indication
of the range in which the state can be accessible. 
We thus show in Fig.~\ref{fig:critical} as a function of $\alpha$ and for different values of $\mu$ the 
threshold $V_c$, the energy splitting $\epsilon_--\epsilon_+$, the difference in the occupation $n_--n_+$,
and in the deflection $f_--f_+$.
As expected the critical voltage $V_c$ decreases as a function of $\alpha$, and in particular, for sufficiently
small $\mu$, it vanishes when $\alpha$ approaches the critical value $\alpha_c$. 
The bound-state energy splitting is monotone in $\alpha$, since the electric field 
increases the localization of the bound state, and thus reduces the difference of the two states.
Its $\alpha$-dependence is rather weak.
Even for $\mu\gg1$ the energy splitting remains of the order of $E_K$, that thus sets the main 
energy scale of the problem.
Quite surprisingly the difference in the fraction of localized charge ($n_--n_+$) 
is not monotonic for small $\mu$ as a function of the electric field.
This is due to the fact that the transition region is approached at different values of $\alpha$ for 
each spin state.
A similar behavior is observed in $f_--f_+$.
One can conclude that the optimal value of $\alpha$ to observe a well defined bound state 
is between 100 and 200. 

{\em Estimates}.
In order now to consider the possibility to observe the two bound states we discuss the typical scales 
of the problem.
Expressing the radius in $nm$ and the length in $\mu$m
$E_K \approx 13.9 (r /L^2)$ mK.
The typical value of $L$ ranges between $0.1$ and $1\mu$m,
leading to a range for $E_K$ between few K to tens of mK,
thus always accessible with standard cryogenics.
The thermal and quantum fluctuations of the displacement of the tip plays also 
an important role, since they define the distinguibility of the displacement 
of the two bound states. 
From \refe{energy} one can write an approximate potential for the tip displacement 
$\delta f= f(1)-f_0$ (with $f_0=n/12$ the equilibrium value): $h_\alpha =2 \alpha (\delta f)^2$.
The equipartition theorem then gives for the thermal fluctuations $\delta f_T=[k_B T/(4\alpha E_K)]^{1/2}$.
Quantum fluctuations -$\delta f_Q$- has the same expression with $k_B T \rightarrow \hbar \omega_m$. 
Since $\hbar \omega_m/E_K=0.0332$ independently of $L$ or $r$ \cite{cleland_foundations_2003,snyman_polarons_2012} then $\delta f_Q= 0.09 /\sqrt{\alpha}$.
Expressing as above $T$ in mK, $L$ in $\mu$m, and $r$ in nm 
$\delta f_T=  0.13 L [T /(r \alpha)]^{1/2}$.
Those values have to be compared with $f_--f_+$ that are at best 0.04.
$f_Q$ is thus 5 times smaller of this value already 
for $\alpha=100$, while in order to keep $\delta f_T$ 
small one needs $T \ll .09 r\alpha/L^2$.
This is realizable for instance choosing $L=.5 \mu$m, $r=2$nm, $\alpha=200$ 
and working at temperatures $T \approx 20$ mK ($E_K$ is 111 mK in this case). 

{ \em Conclusions.} We have shown that combining electrostatic and magnetic gating 
the formation of a spin-polaronic state in a singly clamped CNT becomes possible. 
Electric, magnetic, and mechanical tuning provides an effective manipulation of 
such spin-polaron states offering a controllable magneto-electro-mechanical 
transduction with single electronic charge and spin sensitivity
involving sub nanometer mechanical displacement.

{\em Acknowledgements}
R.S. acknowledges financial support from Idex of Univ. Bordeaux through visiting 
professor program. F.P. acknowledges support from ANR QNM.

\bibliography{SpinPolaron2}

\end{document}